\documentclass[twocolumn,aps,pre,showpacs,floatfix]{revtex4}

\usepackage[dvips]{graphicx}%
\usepackage{dcolumn}
\usepackage{bm}%

\newcommand \be{\begin{equation}}
\newcommand \ee{\end{equation}}
\newcommand \ba{\begin{eqnarray}}
\newcommand \ea{\end{eqnarray}}
\begin{document}
\title
{\Large\bf 
Onset of thermally driven self-motion of a current filament in a bistable semiconductor
structure}

\author{\bf Pavel Rodin}
\email{rodin@mail.ioffe.ru}
\affiliation
{Ioffe Physicotechnical Insitute of Russian Academy of Sciences,\\
Politechnicheskaya 26, 194021, St. Petersburg, Russia}

\setcounter{page}{1}
\date{\today}


\hyphenation{cha-rac-te-ris-tics}
\hyphenation{se-mi-con-duc-tor}
\hyphenation{fluc-tua-tion}
\hyphenation{fi-la-men-ta-tion}
\hyphenation{self--con-sis-tent}
\hyphenation{cor-res-pon-ding}
\hyphenation{con-duc-ti-vi-ti-tes}


\begin{abstract}

We perform an analytical investigation of the bifurcation from
static to traveling current density filaments in a bistable semiconductor
structure with S-shaped current-voltage characteristic. Joule self-heating 
of a semiconductor structure and the effect of temperature on
electron transport are consistently taken into account in
the framework of a generic reaction-diffusion  model with
global coupling.  It is shown that the self-heating is capable to induce
translation instability which leads to spontaneous onset of lateral 
self-motion of the filament along the structure.
This may occur in a wide class of semiconductor structures whose
bistability is caused by impact ionization due to the negative
effect of temperature on the impact ionization rate.
The increment of the translation mode and the instability threshold are
determined analytically.


\end{abstract}
\pacs{05.70.Ln,72.20.Ht,85.30.-z}

\maketitle






\section{Introduction}

Current density filament in a bistable semiconductor
structure with S-shaped current-voltage characteristic
is a high current density domain embedded in a low current density environment 
\cite{ZAK60,RID63,VOL67,BASS70,OSI70,OSI71,OSI73,BONCH75,SCH87}.
Typically filaments appear  spontaneously due to spatial instability of the uniform
current flow when the average current density corresponds to the falling branch 
of the current voltage-characteristic \cite{RID63}.
A single filament survives after the transient (the-winner-takes-all principle)
whereas multifilamentary states are unstable  
\cite{VOL67,BASS70,OSI70,OSI71,OSI73,BONCH75,SCH87}.
A static current filament may undergo further bifurcations related to temporal
and spatial instabilities 
\cite{WAC94,NIE96a,DAT97,BOS00,PEN88,WAC91,NIE92,POG02,POG03,STE03,DEN04,POG04,ROD04}.
Spontaneous onset of the lateral movement is an example of such a bifurcation 
which has been recently studied both experimentally
and theoretically \cite{POG02,POG03,STE03,DEN04,POG04,ROD04}.
This effect is of significant practical importance because
the filament motion delocalizes the Joule heating of a semiconductor structure
operated in the current filamentation regime, and  thus dramatically reduces
the probability of thermal failure \cite{POG03,STE03,DEN04,POG04}. 
Traveling current density filaments bear deep similarities to other 
traveling spots in active media 
\cite{KRI94,ZAI95,BOD97,BOD98,BOD02,MIK02}, in particular, to traveling
wave segments in feed-back controlled light sensitive Belousov-Zabotinsky
reaction \cite{SHO02,SHO02a}. 

The latest exerimental observations of moving
filaments \cite{POG02,POG03,DEN04,POG04} have been made using
a recently developed interferometric mapping 
technique \cite{POG02}. Experiments show that in a device with a stripe-like
geometry the filament travels  in a soliton-like manner with a constant
velocity $\sim 1 \; {\rm m/s }$ and reflects from  the device boundaries.
Both measurements and full-scale numerical simulations of the transport processes
suggest that the filament motion is thermally driven and related  to the
Joule self-heating \cite{POG02,POG03,POG04}. Thermally driven
self-motion is possible due to the negative effect of temperature
on the impact ionization rate \cite{KAY54,SZE}
and may occur in a wide class of  semiconductor structures whose bistability
is caused by impact ionization \cite{WAC91,ROD04}. 
The temperature acts as an inhibitor which provokes translation instability
of a steady filament and causes its self-sustained motion towards a cooler
region \cite{WAC91,ROD04}. 

In this article we present an analytical investigation of the translation 
instability which occurs at the onset of thermally driven filament motion. 
Using a recently suggested three-component reaction-diffusion model for 
current filamentation in presence of the Joule self-heating \cite{ROD04}, 
we perform a stability analysis of a static filament. 
We show that the translation invariance of a static filament in a large structure
breaks in presence of the self-heating. This occurs with
increase of the thermal relaxation time which depends on the heat capacity
of the semiconductor structure and the efficiency of the
external cooling.  The instability threshold  decreases when  
the effect of temperature on the vertical transport becomes stronger
and increases with  heat diffusion coefficient.

\section{The model}

Current filamentation in bistable semiconductors 
can be described by a two-component activator-inhibitor model which
consists of a nonlinear reaction-diffusion equation, accounting for 
the internal activator dynamics in a bistable semiconductor, and
an integro-differential equation, accounting for the applied voltage.
This model, originally developed  for bulk semiconductors with so called 
overheating instability \cite{VOL67,BASS70} and layered semiconductor structures
\cite{OSI70,OSI71,OSI73},  proved to be efficient
for a wide class of bistable semiconductors with various mechanims
of bistability (see  Ref. \onlinecite{SCH01} and references therein). 
We refer to Refs. \onlinecite{WAC95d,ALE98,MEI00} for comprehensive
general formulation of this model.

To decribe the thermally driven motion of a current density
filament, the effect of temperature $T$ on the cathode-anode transport
and  the heat dynamics in the structure should be additionally
taken into account. The respective extension of the conventional
two-component model \cite{WAC95d,ALE98,MEI00} has been recently suggested  
in Ref.\onlinecite{ROD04}: 
\begin{eqnarray}
&&\frac{\partial a}{\partial t} =
\nabla_{\perp} (D_a(a) \nabla_{\perp} a) + f(a,u,T), \label{master} \\
&&\tau_T \, \frac{\partial T}{\partial t} = \ell_T^2 \Delta_{\perp} T +
\bigl( {J u}/{\gamma} +
T_{\rm ext} - T \bigr),
\label{heat}\\
&&\tau_u \frac{d u}{d t} = U_0 - u - R \int_S J(a,u) \, dx dy, \;
\tau_u \equiv RC.
\label{global}
\end{eqnarray}
Here $a$ acts as an activator, $u$ and $T$ as inhibitors.

Eqs.(\ref{master},\ref{global}) are the same as in the conventional 
two-component model \cite{WAC95d,ALE98}. The variable $a(x,y,t)$
characterizes the internal state of the device, $u(t)$ is  the voltage
over the device. The local kinetic function $f$ and the $J(a,u)$
dependence contain the information about the vertical electron transport
in the  cathode--anode direction and together determine the S-shaped 
local current-voltage characteristic (Fig.1). The function $f(a,u)$ has three zeros 
corresponding to the off, intermediate and on states, respectively,
when the voltage $u$ is within the bistability range 
$[u_{\rm h},u_{\rm th}]$ (the upper insert in Fig.1).
The diffusion coefficient $D_a(a)$
characterizes lateral (along the $(x,y)$ plane)
coupling in the spatially extended semiconductor structure. 
Eq.(\ref{global}) is a Kirchhoff's equation for the external
circuit. $R$ is the load resitance switched in series with the semiconductor
structure, $U_0$ is the total applied voltage, $C$ is the effective
capacitance of the sample and the circuit, $S$ is the device area of the 
structure cross-section.
In the following we consider the current-controlled regime. 

Eq.(\ref{heat}) describes 
the heat dynamics in the structure. Similar to the variable $a$,
the temperature $T(x,y,t)$ depends only on the lateral coordinates $x$ and $y$.
$T_{\rm ext}$ is the temperature of 
the external cooling reservoir, $\gamma$ is heat transfer
coefficient, the termal relaxation time $\tau_T$ and diffusion
length $\ell_T$ are given by
\begin{equation}
\tau_T \equiv  c \rho w / \gamma, 
\qquad \ell_T \equiv  \kappa w / \gamma,
\label{thermal}
\end{equation}
where $c$, $\rho$ and $\kappa$ are specific heat, density and heat
conductivity of the semiconductor material, respectively, and 
$w$ is the device thickness in the cathode-anode direction. 

The master equation (\ref{master}) is coupled to the heat equation
(\ref{heat}) via the temperature dependence of the local kinetic
function $f(a,u,T)$. Since the heating suppresses impact ionization, 
the effect of temperature on the vertical transport is negative 
and $\partial_T f < 0$  \cite{footnote1}. 
The direct dependence of the current density $J$ on $T$ is neglected.

The characteristic size of a filament wall is 
$\ell_a \equiv \sqrt{D_a/\partial_a f}$ \cite{WAC95d,ALE98}.
Parameters $\ell_a$ and $\ell_T$ are diffusion lengths of the
activator and inhibitor, respectively.
For simplicity, in the following we take $D_a(a) = \rm {const}$.
Next, we take into account only one  lateral dimension $x$, 
assuming that due to the stripe geometry of the semiconductor
structure ($L_x \gg  \ell_a \gg  L_y$) the current density distribution 
is uniform along the $y$ direction (the lower insert in Fig.1).
We also assume that the filament is located far from the edges of 
the stripe $x= \pm L_x/2$. 
Current density and temperature profiles in a static filament
are shown in the upper panel of Fig.2. For $\ell_a \gg \ell_T$
the temperature profile essentially follows the current density profile
(Fig.2, curve 1 in the upper panel). For $\ell_a \ll \ell_T$
the characteristic size of the hot area is much larger than the filament
size (Fig.2, curve 2 in the upper panel). 

\section{Mechanism of translation instability}

Antisymmetric (with respect to the center of  the filament) fluctuations
of the current density do not  
change the total current and therefore are not suppressed 
by the external circuit in the current-controlled regime.
By increasing the current denisity on one side and decreasing on another
such fluctuation effectively leads to a small shift of a filament, moving
one of the filament edges to a cooler region and another to a hotter region.
Due to the heat inertia of the semiconductor structure the temperature does not follow the current
density instantly. Hence favorable conditions for the impact ionization 
(production of the activator $a$ in terms of Eq.(\ref{master})) appear at the one
of the filament edges, whereas at another edge the impact ionization is suppressed.
This leads to further shift of a filament and, for a sufficiently
long delay of the temperature response, to a self-sustained filament
motion. Experiments  \cite{POG03,POG04} show that in a high quality
structure filaments start to move to the left and to the right with equal
probability. This suggests that motion is triggered by nonequilibrium
fluctuations of the current density.

\section{Setting up the stability problem}

\subsection{Linearization and the eigenvalue problem}

Let us consider a stationary solution $a_0(x)$, $T_0(x)$, $u_0$
of the equations (\ref{master},\ref{heat},\ref{global}):
\begin{eqnarray}
D_a  a_0^{\prime \prime} + f(a_0,u_0,T_0) = 0, \label{master_steady} \\
\ell_T^2 T_0^{\prime \prime} + \frac{J(a_0,u_0) u_0}{\gamma} - T_0 = 0, \label{heat_steady} \\
U_0 - u_0 - R L_y \, \langle J(a_0,u_0) \rangle = 0. \label{global_steady}
\end{eqnarray}
Here the prime $(...)^{\prime}$ denotes the derivative with respect to $x$
and angular brackets $\langle ... \rangle$ denote integration over the
interval $[-L_x/2,L_x/2]$.

Linearization in the vicinity of the stationary solution with respect
to the fluctiations
\begin{eqnarray}
\nonumber
a(x,t) - a_0(x) &=& \delta a(x) \, \exp (\zeta t), \\ \nonumber
u(t) - u_0 &=& \delta u \, \exp (\zeta t), \\   \nonumber
T(x,t) - T_0(x) &=& \delta T(x) \, \exp( \zeta t)  
\end{eqnarray}
leads to the eigenvalue problem
\begin{eqnarray}
\zeta \, \delta a &=& \widehat H_a \delta a + \partial_u f \, \delta u +
\partial_T f \, \delta T , 
\label{master_stab1} \\ \label{heat_stab1}
\tau_T \, \zeta \, \delta T &=&     
\widehat H_T +
\frac{u_0 \partial_a J}{\gamma} \, \delta a +
\frac{J(a_0,u_0)+u_0 \partial_u J}{\gamma} \, \delta u, \qquad \\
\tau_u \, \zeta \, \delta u &=& 
-(1 + R L_y \langle \partial_u J \rangle) \, \delta u -
R L_y \, \langle \partial_a J \, \delta a \rangle.
\label{global_stab1}
\end{eqnarray}
Here
\begin{equation}
\label{operators}
\widehat H_a \equiv D_a \frac{d^2}{d x^2} + \partial_a f,
\qquad
\widehat H_T \equiv \ell_T^2 \frac{d^2}{d x^2} - 1
\end{equation}
and all derivatives are computed at the steady state
$a_0(x)$, $T_0(x)$, $u_0$.

\subsection{The reference case $T=T_{\rm ext}$.}

Let us briefly summarize the results of the stability analysis in
the reference case of constant temperature $T \equiv T_{\rm ext}$  \cite{ALE98}.
In this case  Eq.(\ref{master_stab1}) is reduced to
\begin{equation}
\zeta \, \delta a = \widehat H_a \delta a + \partial_u f \, \delta u. 
\label{master_cold} 
\end{equation}
The eigenfunctions $\Psi_i$, $i=1,2,..$  of the  self-adjoint
operator $\widehat H_a$  correspond to eigenmodes of the stationary
filament in the voltage-controlled regime $u = {\rm const}$ 
(Fig.2, lower panel). These eigenmodes are orthogonal and obey the oscillation
theorem which says that $\Psi_k$ has $(k-1)$ nodes inside the
interval $[-L_x/2,L_x/2]$.  Due to the translation invariance for  
an infinitely large sample the spectrum of $\widehat H_a$ contains
a neutral (so called Goldstone) mode 
\begin{equation}
\Psi_G \sim  a_0^{\prime}
\label{Goldstone1}
\end{equation}
with $\lambda_G = 0$. This mode corresponds to the translation of 
the filament along the $x$ axis. For a solitary filament
$a_0(x)$ has a single maximum. Hence $\Psi_G$ has one node and is identified
as $\Psi_2$ \cite{footnote2}. 
Since $\lambda_2=0$ and $\lambda_1 > \lambda_2$,
the first eigenvalue $\lambda_1$ is positive.
The first eigenmode $\Psi_1$ corresponds to spreading or shrinking
of the current filament. 

Since $a_0$ is symmetric with respect to the center of the filament,
the ``potential term" $\partial_a f$ in the operator $\widehat H_a$
is also symmetric.  Hence $\Psi_1$ and $\Psi_2$
are symmetric and antisymmetric, respectively.
The increase of $\Psi_1$ is prevented by 
the global constraint (\ref{global}) which effectively fixes
the total current when the  external resistance $R$ is sufficiently large
\cite{ALE98}.  In contrast, for $\delta a \equiv \Psi_2$
the last term in the Eq.(\ref{global_stab1}) vanishes 
due to the asymmetry of $\Psi_2$. 
Hence in this case $\delta a$ and $\delta u$ are not coupled,
and we can take $\delta u = 0$. Physically, it happens because $\Psi_2$
does not change the total current and therefore the voltage $u$
also does not change. Thus  for the current-controlled
regime the dynamics of $\Psi_2$ is the same as in
the voltage-controlled regime, and  $\zeta = \lambda_2 = 0$
\cite{VOL67,BASS70,ALE98}.
We conclude that for the two-component  model filaments have neutral
stability with respect to translation when the lateral dimension of the 
semiconductor structure is large ($L_x \gg \ell_a$). 

\subsection{Translation instability}

We shall focus on the second eigenmode $\Psi_2$ with $\lambda_2=0$
and investigate whether for the extended model (\ref{master},\ref{heat},\ref{global})
the respective increment $\zeta$ departs from zero
due to the coupling between the master equation
(\ref{master}) and the heat equation (\ref{heat}). 
As we argued above, in linear approximation the antisymmetric translation
mode does not interact  with external circuit and hence $\delta u = 0$. 
The  set of equations 
(\ref{master_stab1},\ref{heat_stab1},\ref{global_stab1})
reduces to
\begin{eqnarray}
\label{master_stab2}
\zeta \delta a &=& \widehat H_a \delta a + \partial_T f \delta T_G, \\
\tau_T \zeta \delta T &=& \widehat H_T \delta T 
+ \left(u_0 \partial_a J / \gamma \right) \, \delta a.
\label{heat_stab2}
\end{eqnarray}
For the model (\ref{master},\ref{heat},\ref{global})
the translation Goldstone mode $(\delta a_G, \delta T_G )$ has two components 
which are given by
\begin{equation}
\label{Goldstone2}
\delta a_G \equiv a_0^{\prime}, \qquad \delta T_G \equiv T_0^{\prime}.
\end{equation}
By taking the space derivative of Eqs.(\ref{master_steady},\ref{heat_steady})
we find that $(\delta a_G , \delta T_G)$ indeed satisfies 
Eqs.(\ref{master_stab2},\ref{heat_stab2}) for $\zeta=0$:
\begin{eqnarray}
\label{master_stab3}
\widehat H_a \delta a_G + \partial_T f \; \delta T_G = 0, \\
\widehat H_T \delta T_G + \left(u_0 \partial_a J / \gamma \right) \, \delta a_G = 0.
\label{heat_stab3}
\end{eqnarray}
 
\section{Characteristic equation}

The increment $\zeta$ of the translation mode 
can be found from Eqs.(\ref{master_stab2},\ref{heat_stab2}) by means 
of the perturbation theory with respect to the isothermal case
$\zeta = \lambda_2 = 0$.
Substituting $\delta a = \delta a_G$ in Eq.(\ref{heat_stab2})
we obtain
\begin{equation}
\label{heatresponce}
\tau_T \, \zeta \, \delta T =
\widehat H_T \delta T + \left( u_0 \partial_a J / \gamma \right) \; \delta a_G.
\end{equation}
The solution $\delta T \left[ \zeta, \delta a_G  \right]$ determines
the temperature response to the variation $\delta a_G$ 
of the activator distribution and depends 
on both $\delta a_G$ and $\zeta$.  Substitution of 
$\delta T \left[ \zeta, \delta a_G \right]$ into Eq.(\ref{master_stab2})
leads to 
\begin{equation}
\label{INT1}
\zeta \, \delta a_G = \widehat H_a \, \delta a_G 
+ \partial_T f \; \delta T [\zeta, \delta a_G].
\end{equation} 
Now note that according to Eqs.(\ref{heat_stab3},\ref{heatresponce}) 
$\delta T_G$
can be presented as 
$$
\delta T_G \equiv \delta T \left[\zeta =0, \delta a_G \right].
$$
Therefore according to Eq.(\ref{master_stab3}) 
\begin{equation}
\label{INT2}
\widehat H_a \delta a_G + \partial_T f \; \delta T [\zeta=0, \delta a_G]=0.
\end{equation} 
Substracting (\ref{INT2}) from (\ref{INT1}), muliplying by $\delta a_G$
and  integrating over $x$, we come to the equation for $\zeta$:
\begin{equation}
\zeta = \frac
{\langle \partial_T f 
\left(\delta T[\zeta, \delta a_G]-\delta T[\zeta=0, \delta a_G]\right)
a_0^{\prime} \rangle}
{\langle (a_0^{\prime})^2 \rangle}.
\label{characteristic0}
\end{equation}

In the next two sections we find $\delta T[\zeta, \delta a_G]$ and 
solve the equation (\ref{characteristic0}) for
the limiting cases $\ell_a \gg \ell_T$ and $\ell_a \ll \ell_T$.
  
\section{Weak heat diffusion}

In this section we consider the case $\ell_a \gg \ell_T$
when lateral diffusion of the 
internal parameter $a$ is much more efficient than the heat diffusion.
In this case the temperature profile essentially
follows the current density profile (Fig.2, curve 1 in the upper panel). 
In the limiting case $\ell_T \rightarrow 0$ the solution
of Eq.(\ref{heatresponce}) is given by
\begin{equation}
\delta T[\zeta,\delta a_G] = \frac{u_{0} \, \partial_a J}{\gamma(\tau_T \zeta + 1)} \; \delta a_G.
\label{heatresponce1}
\end{equation}
With this input Eq.(\ref{characteristic0}) yields
\begin{equation}
\label{characteristic1}
\zeta = \frac{\zeta}{\zeta + \tau_T^{- 1}} \;
\Lambda, \qquad 
\Lambda \equiv - \frac{u}{\gamma} 
\frac{ \langle \, \partial_T f  \; \partial_a J \; ( a_0^{\prime})^2
\rangle} { \langle (a_0^{\prime})^2 \rangle}.
\end{equation}
Two solutions $\zeta = 0$ and $\zeta = \Lambda - \tau_T^{-1}$
of this equation
appear to intersect at $\tau_T = \Lambda^{-1}$ in the degenerate case
($\lambda_G = 0$) under consideration. The relevant branch can be chosen
by starting from  $\lambda_G \ne 0$ and taking the limit 
$\lambda_G \rightarrow 0$ (see Appendix).
This leads to the piecewise linear dependence:
\begin{eqnarray}
\label{increments1}
\zeta = 0 \qquad &{\rm for}& \qquad \tau_T  < \Lambda^{-1} ;\\
\zeta = \Lambda - \tau_T^{-1} \qquad &{\rm for}& \qquad \tau_T >
\Lambda^{-1}.
\nonumber
\end{eqnarray}
According to (\ref{increments1}), for sufficiently small $\tau_T$
the filament remains neutral with respect to translation.
The bifurcation from static to traveling filaments occurs with increase
of $\tau_T$ at $\tau_T = \Lambda^{-1}$ .

Note that in the limiting case $\tau_T$ = 0 the temperature
$T$ instantly follows all changes of the current density.
In this case $T$ is a local function of the parameter $a$
and can be eliminated by re-defining the local kinetic function according to
\begin{equation}
\widetilde f(a,u) \equiv f \left( a, u, T = T_{\rm ext}
+ \frac{J(a,u) \; u}{\gamma} \right).
\label{tilde_f}
\end{equation}
In this way the three-component model is reduced back to the two-component model
which with respect to the filament stability  is equivalent to the isothermal model
for $T \equiv T_{\rm ext}$. Hence for $\tau_T \rightarrow 0$ filament has
neutral stability, in agreement with Eq.(\ref{increments1}). 
In the opposite case of slow self-heating ($\tau_T = \infty$)
the increment $\zeta$ has the largest value 
$\zeta =\Lambda$. Generally, the heat diffusion smoothes the temperature response
$\delta T \left[\zeta, \delta a_G \right]$,
thus shifting the instability threshold to larger values of $\tau_T$.

\section{Strong heat diffusion}

In the opposite limiting case $\ell_T \gg \ell_a$
the lateral spreading of heat is more efficient than
the lateral spreading of the current density
(Fig.2, curve 2 in the upper pannel). We restrict the analysis to a 
filament with a flat top (Fig.3). Such filaments are typical for large 
structures ($L_x \gg l_a$). The filament width and characteristic
width of the filament wall are denoted as $W$ and  $\ell_f$, respectively.
Note that $\ell_f \sim \ell_a$ \cite{WAC95d,ALE98}.
 
In this case the translation mode  $\Psi_G \sim a_0^{\prime}$ is distinct
from zero only within the filament walls where it has a characteristic value
$\pm (a_{on} - a_{off})/\ell_f$ (Fig.3, lower panel). 
Taking into account the scale separation
$\ell_f \ll \ell_T$, we can approximate $\delta a_G$ as
\begin{equation}
\delta a_G = \sqrt{ \ell_f / 2 } \;
\left[ \delta(-W/2)-\delta(W/2)\right],
\label{translation}
\end{equation}
where $\delta(x)$ is the delta-function, the center of the filament
is taken at $x=0$ and the prefactor $\sqrt{\ell_f/2}$ provides 
normalization $\langle (\delta a_G)^2 \rangle = 1$.
For $\delta a_G$ given by (\ref{translation}) the 
solution $\delta T \left[\zeta, \delta a_G \right]$
of Eq.(\ref{heatresponce}) is
\begin{eqnarray}
\label{GreenFunc}
\delta T &=&
\frac{M \widetilde \ell_T}{2}
\left[1 - \exp\left[-\frac{W}{\widetilde \ell_T}\right]\right]
\exp\left(\frac{x+(W/2)}{\widetilde \ell_T}\right), \\ \nonumber
&{\rm for}& \qquad x < - W/2, \\ \nonumber
\delta T &=&
- M \widetilde \ell_T
\exp \left[- \frac{W}{2 \widetilde \ell_T} \right]
\sinh \left(\frac{x}{\widetilde \ell_T} \right), \\ \nonumber
&{\rm for}& \qquad -W/2 < x < W/2, \\ \nonumber
\delta T &=&
- \frac{M \widetilde \ell_T}{2}
\left[1 - \exp \left[- \frac{W} {\widetilde \ell_T} \right]\right]
\exp\left(-\frac{x - (W/2)}{\widetilde \ell_T}\right) \\ \nonumber
&{\rm for}& \qquad x > W/2,
\end{eqnarray} 
where
\begin{eqnarray}
\label{notation}
M(\zeta) \equiv \sqrt{\frac{\ell_f}{2}}
\frac{u_0 \, \partial_a J}{\ell_T^2 \, \gamma \, (1 + \tau_T \zeta)},
\qquad
\widetilde \ell_T \equiv \frac{\ell_T}{\sqrt{1 + \tau_T \zeta}}. 
\end{eqnarray}

Substituting $\delta T[\zeta, \Psi_G)]$ into (\ref{characteristic0})
we obtain 
\begin{eqnarray}
\label{characteristic2}
\zeta = 
\partial_T f \, \sqrt{\frac {\ell_f}{2}} \, 
\left( M(\zeta) \, \widetilde \ell_T(\zeta) \,
\left[1 - \exp\left(-\frac{W}{\widetilde \ell_T(\zeta)}\right)\right]
\right.
\nonumber \\
\left. - M(0) \, \ell_T \,
\left[1 - \exp\left(-\frac{W}{\ell_T}\right)\right]
\right).
\end{eqnarray}
Taking into account that near the bifurcation point $\zeta$ is
small, we expand the right handside of Eq.(\ref{characteristic2}) 
over ($\tau_T \zeta$) and come to the equation
\begin{eqnarray}
\label{characteristic3}
\zeta = \Omega \, \frac{\tau_T \, \zeta}{2 + \tau_T \, \zeta}, 
\qquad
\Omega \equiv \frac{\ell_f}{\ell_T} 
\frac{u_0 \; \partial_T f \; \partial_a J}{2 \gamma}
A(W), \\ \nonumber
A(W) \equiv 1 - \left(1 + \frac{W}{\ell_T}\right)
\exp \left(-\frac{W}{\ell_T}\right).
\end{eqnarray}
Eq.(\ref{characteristic3}) coincides with Eq.(\ref{characteristic1})
when $\Lambda$ is replaced by $\Omega$ and $\tau_T$ by
$(\tau_T/2)$. Hence the solutions have the same structure as in
Eq.(\ref{increments1}):
\begin{eqnarray}
\label{increments2}
\zeta = 0 \qquad &{\rm for}& \qquad \tau_T  < 2 \Omega^{-1}, \\
\zeta = \Omega - 2 \tau_T^{-1} \qquad &{\rm for}& \qquad \tau_T >
2 \Omega^{-1}.
\nonumber
\end{eqnarray}

\section{Discussion}

\subsection{Onset of filament motion for weak and strong heat diffusion}

For $\ell_T \ll \ell_a$ the condition for the onset of filament motion
is given by Eq.(\ref{increments1}) and can be presented as
\begin{equation}
\label{stab1}
\tau_T > \tau^{\rm th}_1 \equiv \Lambda^{-1} = 
\frac{\gamma}{u_0} 
\frac{\langle(a_0^{\prime})^2 \rangle}
{\langle \partial_T f \; \partial_a J \; (a_0^{\prime})^2 \rangle}.
\end{equation}
In the opposite limiting case $\ell_T \gg \ell_a$
this condition is given by
\begin{equation}
\label{stab2}
\tau_T > \tau^{\rm th}_2 \equiv \frac{4 \ell_T}{\ell_f A(W)}
\frac{\gamma}{u_0 \; \partial_T f \; \partial_a J}.
\end{equation}
In both cases $\ell_a \ll \ell_T$
and $\ell_a \gg \ell_T$ the instability
occurs when the thermal relaxation time $\tau_T$ is sufficiently
large. This suggests that $\tau_T$ is a bifurcation
parameter also in the intermediate case $\ell_a \sim \ell_T$.

The relation between $\tau^{\rm th}_1$ and  $\tau^{\rm th}_2$
becomes transparent when $\partial_T f$ and 
$\partial_a J$ are considered as constants.
(Note also that $\ell_f \sim \ell_a$.)
In this case
\begin{equation}
\label{relation}
\tau^{\rm th}_2 \sim \frac{4 \ell_T}{\ell_a  A(W)} \,  \tau^{\rm th}_1.
\end{equation}
For wide ($W \gg \ell_T$)  and 
narrow ($W \ll \ell_T$) filaments the 
expansion of $A(W)$ over $W/\ell_T$ (see Eq.(\ref{characteristic3}))
leads to
\begin{eqnarray}
\label{times2}
\tau^{\rm th}_2 \sim \frac{\ell_T}{\ell_a } \,  \tau^{\rm th}_1
\qquad {\rm for} \qquad W \gg  \ell_T, \\ \nonumber
\tau^{\rm th}_2 \sim \frac{2 (\ell_T)^3}{\ell_a W^2 } \,  \tau^{\rm
th}_1 \qquad {\rm for} \qquad W \ll  \ell_T.
\end{eqnarray}
Thus the threshold time $\tau_2^{\rm th}$ exceeds $\tau_1^{\rm th}$ 
by a factor which is always larger than $\ell_T/\ell_a$.

For $\ell_T \gg \ell_a$ the instability threshold 
has been previously obtained in Ref.\onlinecite{ROD04}  
for the current density profile as shown in Fig.3.
The argument  is based on the finding that the velocity $v$ 
of the stationary filament motion is proportional
to the difference of temperatures  at the front and back filament edges 
$\Delta T$: $v = C \Delta T$ \cite{ROD04}. This temperature difference, in turn,
depends on the $v$. Hence the velocity of stationary motion
can be found from the equation $v=C \Delta T(v)$. Since for $v=0$
the temperature profile is symmetric, $\Delta T(0)=0$.
Therefore $v=0$ is always a solution of this equation.
However, this solution is unstable if  $\Delta T(v)$ increases
faster than $v/C$ with increase of $v$. In this case 
when $v$ deviates  from $v=0$, the "overproduction" of the 
the temperature difference $\Delta T$ leads to further increase of $v$.
The respective condition is given by \cite{ROD04}
\begin{eqnarray}
\label{stab3}
&&\frac{v_0}{2 v_T} A(W) > 1,  \;
v_0 \equiv 
\frac
{(J_{\rm on}-J_{\rm off})u_{0}}
{\gamma \; \langle (a_0^{\prime})^2 \rangle}
\int_{a_{\rm off}}^{a_{\rm on}} \partial_T f \; da. \qquad
\end{eqnarray}
Here $v_0$ has the meaning of the upper limit of the filament velocity 
and $v_T \equiv \ell_T / \tau_T$ is the thermal velocity.
$J_{\rm on}$ , $J_{\rm off}$
and $a_{\rm off}$, $a_{\rm on}$ are  the current denisties
and the values of the variable $a$ correspoding to
the uniform on and off states  at $u=u_0$, respectively (Fig.3).
$A(W)$ is exactly the same as in Eq.(\ref{characteristic2}).
The criterium (\ref{stab3}) can be re-written as
\begin{equation}
\label{stab4}
\tau_T > \widetilde \tau^{\rm th}_2,
\qquad \widetilde \tau^{\rm th}_2 = 
\frac{2}{A(W)} \frac{\ell_T}{v_0}.
\end{equation}
Taking into account that for a wide filament
\begin{equation}
\label{estimate}
\langle (a_0^{\prime})^2 \rangle \approx 2
(a_{\rm on} - a_{\rm off})^2/ \ell_f
\end{equation}
and considering the function 
$\partial_T f$ as a constant, we find
\begin{equation}
\label{v_0}
v_0 \approx - \frac{\partial_T f \, u_0}{2 \gamma}
\frac{J_{\rm on} - J_{\rm off}}{a_{\rm on} - a_{\rm off}}.
\end{equation}
Hence
\begin{equation}
\widetilde \tau^{\rm th}_2 = \frac{4 \ell_T}{ \ell_f A(W)}
\frac{\gamma}{u_0 \;  \partial_T f}
\frac{a_{\rm on} - a_{\rm off}}{J_{\rm on} - J_{\rm off}}.
\label{stab5}
\end{equation}
Comparing Eq.(\ref{stab2}) and Eq.(\ref{stab5}),
we see that $\widetilde \tau^{\rm th}_2$ coincides with
$\tau^{\rm th}_2$ (see Eq.(\ref{stab2}) if we approximate 
the partial derivative $\partial_a J$ by the finite difference
$$
\partial_a J \sim 
(J_{\rm on} - J_{\rm off})/(a_{\rm on} - a_{\rm off}).
$$

\subsection{Traveling filament}

Remarkably, the difference between the regimes of weak
and strong heat diffusion, which is well-pronounced near
the bifurcation from static to traveling filament, vanishes for 
fast self-sustained filament motion with velocity $v \gg v_T$.
In this regime the static thermal diffusion
length $\ell_T = \sqrt{D_T \tau_T}$ should be replaced by
${\cal L}_T = \sqrt{D_T \tau_f}$, where $\tau_f = W/v$
is the time of the filament passage over its width $W$ and 
$D_T =  \kappa/c \rho$ is thermal diffusivity. For 
${\cal L}_T \ll W$ the heat diffusion is negligible.
This condition yields
\begin{equation}
\label{conditionsqrt}
v \gg  D_T/W = v_T  \ell_T  / W.
\end{equation}
For typical filament width $W \sim 10 \, {\rm \mu m}$
the condition (\ref{conditionsqrt}) yields $v \gg 10^3 \; {\rm cm/s}$ for Si 
and $v \gg 250 \; {\rm cm/s}$ for GaAs 
($D_T = 0.92$ and $0.25 \; {\rm cm^2/s}$, respectively).

If the condition ${\cal L}_T \ll W$ is met togeter with
the condition  $\tau_f \ll \tau_T$, the heating is not only local but also adiabatic.
(Note that for $\ell_T \sim W$ these two conditions are equivalent.)
In this case the rate of temperature increase in each point is proportional
to the local power dissipation. The velocity of such fast narrow filament
has been obtained in Ref.\onlinecite{ROD04}:
\begin{equation}
\label{sqrt1}
v \approx \sqrt{v_0 \, W / \tau_T}.
\end{equation}
Substituting the definitions $\tau_T$ and $v_0$ 
(see Eqs.(\ref{thermal},\ref{v_0})),
we find that the filament velocity is proportional to the square root
of the total Joule power which is dissipated in the filament:
\begin{equation}
\label{sqrt2}
v = \sqrt{\frac{K \cdot P}{c \rho}}, \;
P \equiv \ \frac{(J_{\rm on} - J_{\rm off}) W u_0}{w}
\approx \frac{J_{\rm on} W  u_0}{w},
\end{equation}
where $c \rho$ is specific heat per unit volume and the coefficient
$K$ is given by
\begin{equation}
\label{coefficient}
K \equiv \frac{1}{\langle (a_0^{\prime})^2 \rangle}
\int_{a_{\rm off}}^{a_{\rm on}} \partial_T f da 
\sim \frac{\ell_f \; \partial_T f}{2 (a_{\rm on} - a_{\rm off})^2}.
\end{equation}
Alternative elementary derivation of the square root dependence is
presented in Ref.\onlinecite{POG04}. 
 
Apparently, the square root dependences (\ref{sqrt1},\ref{sqrt2})
reflect the experimental situation which is favorable for
observation of traveling filaments because large velocity corresponds
to low instability threshold. It has been reported 
\cite{POG04} that the results of experimental measurements 
of the filament velocity in a Si device 
($v \sim 3 \cdot 10^4 \, {\rm cm/s}$, $W \sim 10 \, {\rm  \mu m}$)
indeed obey the square root dependence.

\subsection{The type of bifurcation}

The piecewise linear dependence (\ref{increments1},\ref{increments2})
of the increment $\zeta$ on the bifurcation parameter $\tau_T^{-1}$ 
as well as  the square root behaviour of the filament velocity near the
bifurcation point, predicted in Ref.\onlinecite{ROD04}, are qualitatively
the same as obtained  in Ref.\onlinecite{BOD98} for the three-component 
``cubic" model for a traveling spot. These features seem to be typical
for a degenerate ``drift pitchfork" bifurcation associated with $\lambda_G = 0$.
Note that the mechanism of self-motion described here for one-dimensional
spatial domains is also valid for two-dimensional domains. 
The difference between the model (\ref{master},\ref{heat},\ref{global})
and three-component models for traveling spots studied in
Refs.\cite{KRI94,ZAI95,BOD97,BOD98,BOD02} has been  discussed in 
Ref.\onlinecite{ROD04}.

\subsection{The alternative mechanism for self-motion of a current filament}

Generally, the self-motion of a current filament becomes possible when 
a semiconductor  structure, apart from the activator mechanism
which leads to bistability, posseses an internal mechanism of slow inhibition.
As we described here, such mechanim can appear due to the effect of Joule heating
on the cathode-anode electron transport. Another inhibitor mechanism
of purely electrical origin has been discussed in Refs.\cite{NIE92,NIE95,NIE97,NIE04}.
For this mechanism inhibition is associated with the internal voltage drop across 
the plasma layer inside of the semiconductor structure. The regions
of activation and inhibition are spatially separated.
The effective reaction-diffusion model \cite{NIE92,RAD87,KER94} for such two-layer
system has the same structure as (\ref{master},\ref{heat},\ref{global}). 
Recent numerical simulations demonstrate that this mechanism leads to filament
motion during the switching-off transients in power diodes \cite{NIE04}.
The applicability of this concept to thyristor-like 
$p^{+}-n^{+}-p-n^{-}-n^{+}$ structures discussed in \cite{NIE92} should be considered
in view of the latest studies \cite{WIE95,GOR96,GOR9902} which reveal complexity
of the vertical transport processes in these multilayer structures.

\section{Conclusions}

Joule self-heating of a static current-density filament may
lead to the translation instability and the onset of thermally driven
self-motion. This may occur in a wide class of semiconductor 
devices whose bistability is caused by impact ionization because 
impact ionization coefficients decrease with temperature. 
The eigenvalue of the respective translation mode,
which is zero when heating is neglected, becomes positive
due to the suppressive effect of temperature on the electron transport
in the cathode-anode direction. Increments of the translation mode are
found for  the limiting cases of weak (\ref{increments1}) and 
strong (\ref{increments2}) heat diffusion. In both cases
the instability threshold (\ref{stab1},\ref{stab2}) is controlled
by the thermal relaxation time $\tau_T$ and is directly proportional 
to $\partial_T f$. It scales as $\ell_T/\ell_a$ with increase 
of the thermal diffusion length $\ell_T$.  

    \begin{acknowledgments}
The author is grateful to A. Alekseev for critical reading 
of the manuscript and the hospitality at the 
mathematical department of the University of Geneva
and to D. Pogany for helpful discussions.
This work has been supported by the Swiss National 
Science Foundation.
    \end{acknowledgments}

\appendix*

\section{increment of the translation mode}

To indentify the physical branch of the solution of 
Eq.(\ref{characteristic1}), we note that  
for  $\lambda_G \ne 0$ this equation becomes
\begin{equation}
\label{appendix1}
\zeta = \lambda_G + \frac{\zeta}{\zeta + \tau_T^{-1}} \; \Lambda.
\end{equation}
Smooth solution of this quadratic equation
\begin{equation}
\label{appendix2}
\zeta = \frac{\lambda_G + \Lambda - \tau_T^{-1}}{2}+
\sqrt{\frac{(\lambda_G + \Lambda - \tau_T^{-1})^2}{4} + \frac{\lambda_G}{\tau_T}}
\end{equation}
for $\lambda_G \rightarrow 0$ tends to the piecewise linear 
dependence given by Eq.(\ref{increments1}). 
Note the similarity of Eq.(\ref{appendix1},\ref{appendix2}) and Eq.(9)
in Ref.\onlinecite{BOD97}.


\newpage

\begin{figure*}
\begin{center}
\includegraphics*[width=7.5cm,height=10cm,angle=270]{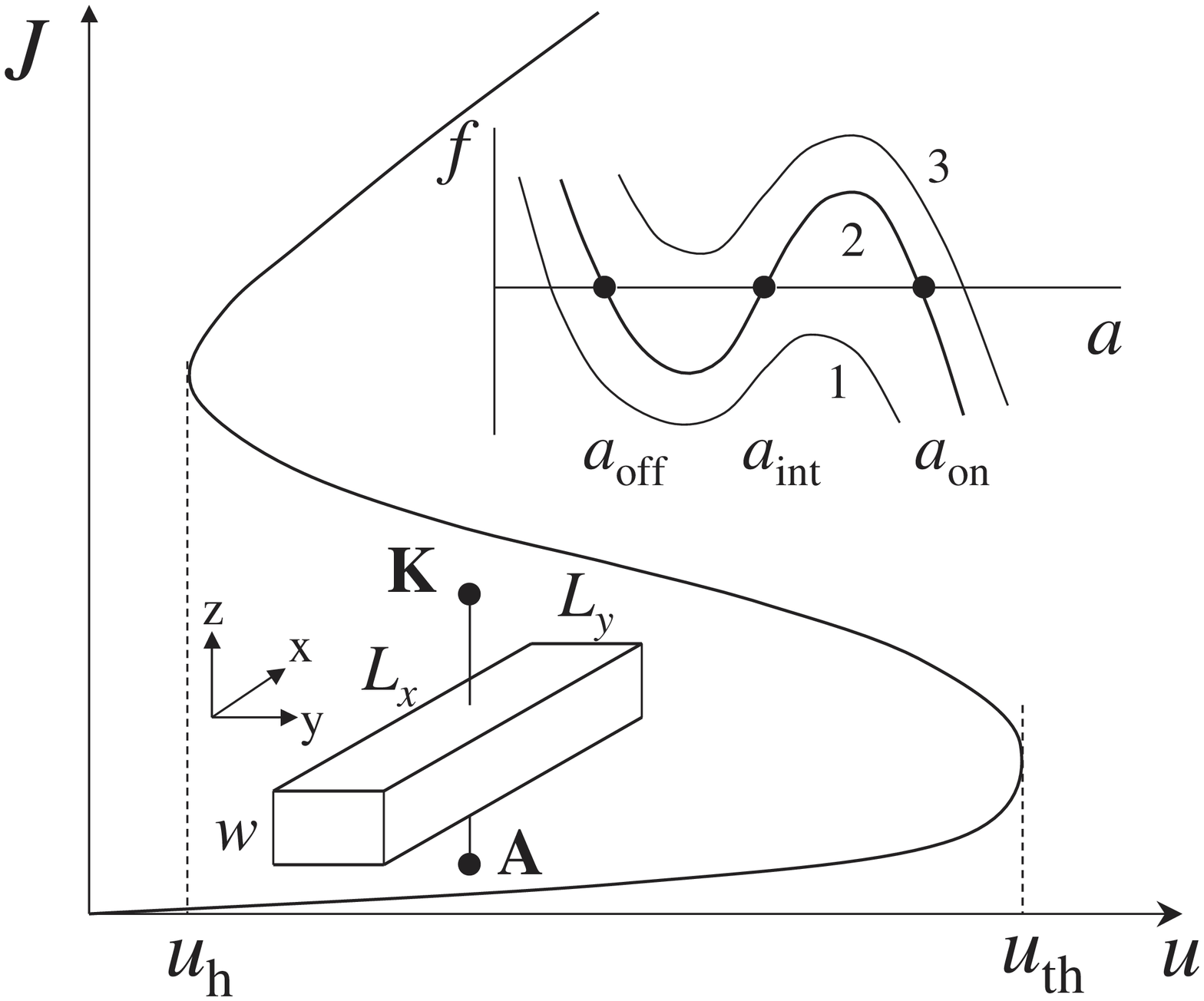}
\end{center}
\caption{
S-shaped current-voltage characteristic $J(u)$ of a bistable
semiconductor structure. The upper insert shows the local
kinetic function $f(a,u)$ for $u < u_{\rm h}$,
$u_{\rm h} < u < u_{\rm th}$ and $u > u_{\rm th}$
(curves 1,2 and 3, respectively),
where $u_{\rm h}$ and $u_{\rm th}$ are the hold and the threshold voltages
at the edges of the bistability range. For $u_{\rm h} < u < u_{\rm th}$
the local kinetic function has three zeros corresponding the off,
intermediate and on branches of the current-voltage characterisics,
respectively.
The lower insert sketches the rectundular semiconductor structure
elongated along the $x$ axis.
}
\label{Fig1}
\end{figure*}

\begin{figure*}
\begin{center}
\includegraphics*[width=7.5cm,height=10cm,angle=270]{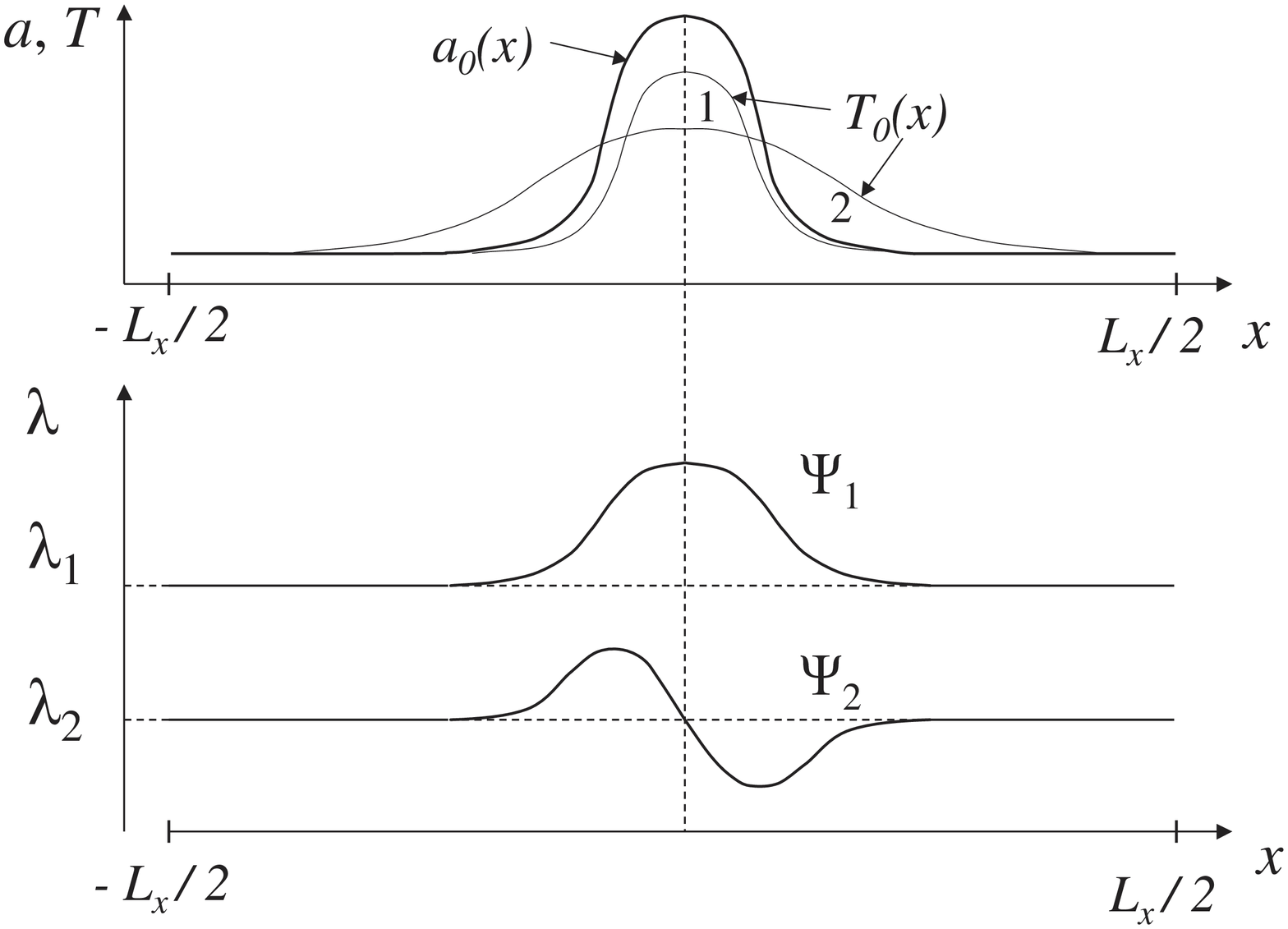}
\end{center}
\caption{
Profile $a_0(x)$ of the activator and the temperature profile
$T_0(x)$ in a steady filament (the upper panel). 
The current denisity profile $J(a_0(x),u)$ is quailitatively
the same as $a_0(x)$. The temperature
profile $T_0(x)$ is sketched for weak ($\ell_T \ll \ell_a$, curve 1)
and strong ($\ell_T \gg \ell_a$, curve 2) heat diffusion.
Note that in our analysis $L_x$ is much larger than the filament width.
The lower panel shows two first eigenfunctions $\Psi_1$ and
$\Psi_2$ of the operator $\widehat H_a$ and their eigenvalues
$\lambda_1$, $\lambda_2$. $\Psi_1$ and $\Psi_2$ are eigenmodes
of the steady filament in a voltage controlled regime and correspond
to expansion(shrinking) and translation of the filament, 
respectively. In the isotheramal case $\lambda_2 = 0$ and 
$\Psi_2 \equiv \Psi_G \sim a_0^{\prime}$.
}
\label{Fig2}
\end{figure*}

\begin{figure*}
\begin{center}
\includegraphics*[width=7.5cm,height=10cm,angle=270]{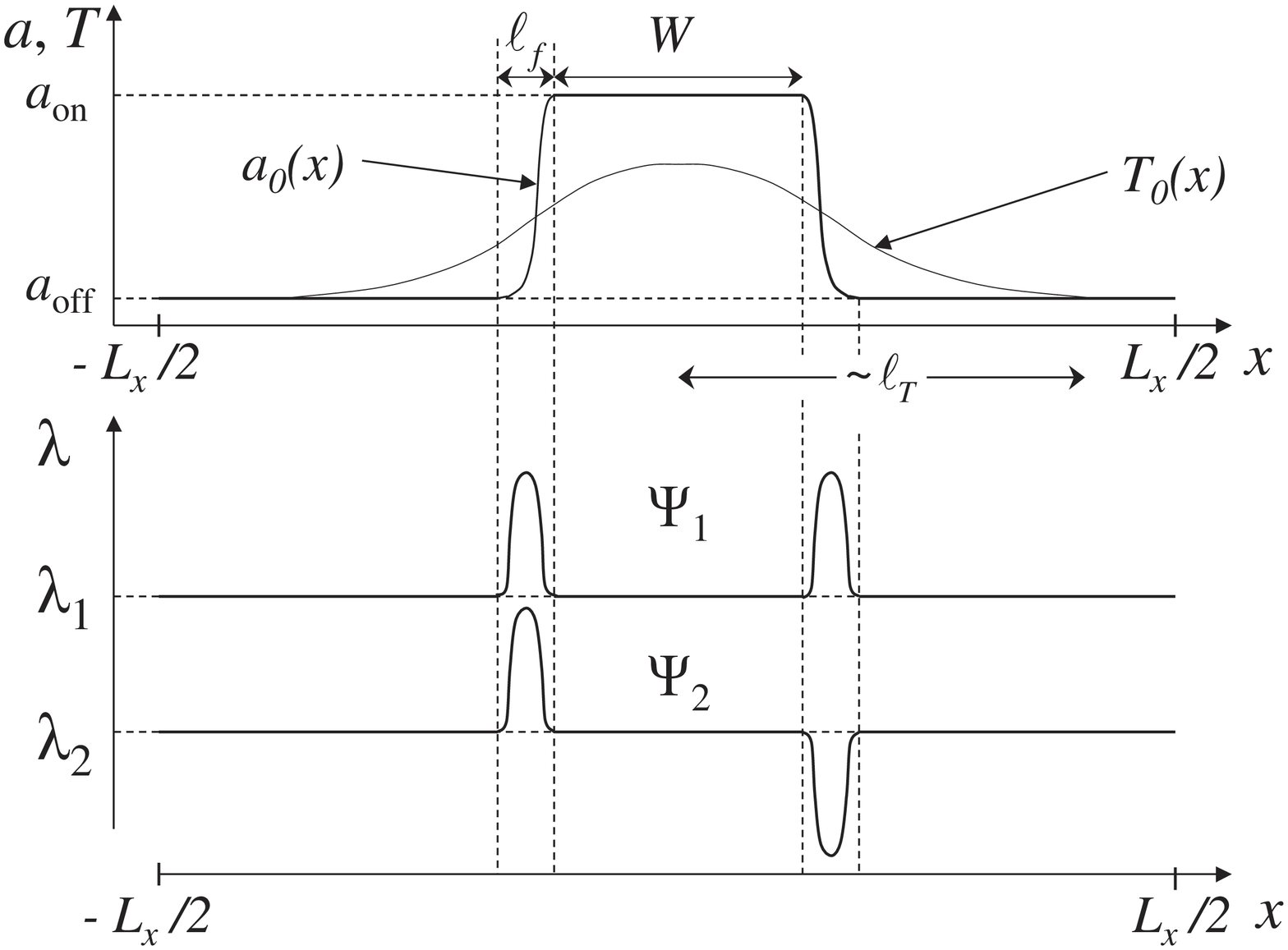}
\end{center}
\caption{
Profile $a_0(x)$ of the activator and the temperature
profile $T_0(x)$ in a wide steady filament (the upper panel).
The filament width $W$, the width of the filament wall $\ell_f$  
and the thermal diffusion length $\ell_T$ are indicated. Note that
in our analysis $L_x \gg W \gg \ell_f$ and $\ell_T \gg \ell_f$, whereas 
the $\ell_T/W$ ratio is arbitrary. As in Fig.2, the low panel sketches 
two first eigenmodes $\Psi_1$ and $\Psi_2$ of the operator $\widehat H_a$ 
and their eigenvalues $\lambda_1$,$\lambda_2$. 
}
\label{Fig3}
\end{figure*}


\begin{thebibliography}{99}
\bibitem{footnote2}
Note that for Neumann boundary conditions imposed 
on the activator $a$, the functions 
$\Psi_2$ and $\Psi_G$  satisfy Neumann and Dirichlet
boundary conditions, respectively, 
and therefore do not exactly coincide \cite{ALE98}. This difference
becomes important  when $L_x$ is comparable to the filament size.
Then the eigenvalue $\lambda_2$ becomes positive refelecting
the attractive effect of boundaries on the filament.
This effect decreases exponentially with $L_x$
and is not considered here.
\end{thebibliography}
\end{document}